\documentstyle[aas2pp4]{article}
\begin{document}
\title{I ZW 18 -- A NEW WOLF-RAYET GALAXY\altaffilmark{1}}
\author{Yuri I. Izotov\altaffilmark{2}}
\affil{Main Astronomical Observatory, Ukrainian National Academy of Sciences,
Goloseevo, Kiev 252650, Ukraine \\ Electronic mail: izotov@mao.kiev.ua}
\author{Craig B. Foltz}
\affil{Multiple Mirror Telescope Observatory, University of Arizona,
Tucson, AZ 85721 \\ Electronic mail: cfoltz@as.arizona.edu}
\author{Richard F. Green}
\affil{National Optical Astronomy Observatories,
Tucson, AZ 85726 \\ Electronic mail: rgreen@noao.edu}
\author{Natalia G. Guseva}
\affil{Main Astronomical Observatory, Ukrainian National Academy of Sciences,
Goloseevo, Kiev 252650, Ukraine \\ Electronic mail: guseva@mao.kiev.ua}
\and
\author{Trinh X. Thuan}
\affil{Astronomy Department, University of Virginia, Charlottesville,
VA 22903 \\ Electronic mail: txt@virginia.edu}

\altaffiltext{1}{Spectroscopic observations presented herein were obtained
with the Multiple Mirror Telescope, a facility operated jointly by the
Smithsonian Institution and the University of Arizona.}
\altaffiltext{2}{Visiting astronomer, 
Kitt Peak National Observatory, National Optical Astronomy Observatories,
operated by the Association of Universities for Research in Astronomy,
Inc., under cooperative agreement with the National Science Foundation.}

\begin{abstract}

We report the discovery of broad Wolf-Rayet emission lines in the 
{\sl Multiple Mirror Telescope (MMT)} spectrum of the NW component of I Zw 18,
the 
lowest-metallicity blue compact dwarf (BCD) galaxy known. Two broad
WR bumps at the wavelengths $\lambda$4650 and $\lambda$5800 are detected
indicating the presence of WN and WC stars. The total numbers of WN and WC
stars inferred from the luminosities of the broad He II $\lambda$4686 and
C IV $\lambda$5808 lines are equal to 17$\pm$4 and 5$\pm$2, respectively. The WR-to-O
stars number ratio is equal to $\sim$0.02, in satisfactory agreement with
the value predicted by massive stellar evolution models with enhanced mass 
loss rates. The WC stars in the NW component of I Zw 18
can be responsible for the presence of the nebular He II $\lambda$4686 emission
line, however the observed intensity of this line is several times larger
than model predictions and other sources of ionizing radiation at
wavelengths shorter than 228\AA\ are necessary.
\end{abstract}


\section {INTRODUCTION}

   The presence of large numbers of Wolf-Rayet (WR) stars in 
star-forming galaxies is well established (Allen, Wright \& Goss 1976;
D'Odorico \& Rosa 1981; Osterbrock \& Cohen 1982; 
D'Odorico, Rosa \& Wampler 1983; Hutsemekers \& Surdej 1984;
Kunth \& Joubert 1985; Kunth \& Schild 1986; Sargent \& Filippenko 1991;
Conti 1991; Vacca \& Conti 1992). These galaxies are often called
WR-galaxies and they are of quite heterogeneous types. 
We focus here on the problem
of detection of WR stars in the low-mass and low-metallicity blue compact
dwarf (BCD) galaxies. Systematic spectroscopic studies of BCDs have shown
that in the spectra of $\sim$1/3 of BCDs, broad WR bumps
characteristic of late WN stars are present, mainly
at $\lambda$4650
(Izotov, Thuan \& Lipovetsky 1994, 1997; Izotov \& Thuan 1997a). 
The intensity of these bumps
decreases with decreasing metallicity, in agreement with predictions of
massive star evolution models and models of evolutionary
population synthesis for star-forming regions (Maeder \& Meynet 1994;
Cervi\~{n}o \& Mas-Hesse 1994; Leitherer \& Heckman 1995; Meynet 1995; Schaerer 
1996). The lowest metallicity BCD
in which WR stars have been detected has $\sim$$Z_\odot$/10,
although BCDs can be as metal-deficient as 
$Z_\odot$/50. Are WR stars present in these extremely metal-deficient BCDs?
In principle, massive stellar evolution theory (e.g. Maeder \& Meynet
1994) does predict the evolution of the most massive low-metallicity stars 
through
the WR stage. However, since the efficiency of 
mass loss by stellar wind decreases
with decreasing metallicity, the total number of WR stars and the total 
duration
of the WR phase in a star formation episode 
are significantly reduced at low metallicities.
This trend led Schaerer et al. (1997) to conclude that in some metal-deficient 
BCDs, weak WR spectral
features are not detected simply because of inadequate signal-to-noise ratio.

Several recent observations of low-metallicity BCDs have 
suggested that massive stars with mass loss are indeed present in 
galaxies with heavy element abundances less than $Z_\odot$/20.
Imagery of I Zw 18 with {\sl HST} by Hunter \& 
Thronson (1995) has resolved its NW and SE components  
into stars, with the brightest star having $V$ $\sim$ 22 mag. Those authors 
attempted to find WR stars using a narrow-band image
in the He II $\lambda$4686 line and  detected only two marginal WR candidates.
They expected a large population of WR stars,
 given the presence of a large number of red supergiants, so concluded that
I Zw 18 is not a BCD  with many Wolf-Rayet stars.
Recently, Izotov \& Thuan (1997b), from 4m Kitt Peak Mayall telescope
spectrophotometry of I Zw 18, noted that WC stars are possibly present in the
NW component. Finally, Thuan \& Izotov (1997) have found evidence for stars 
with mass loss through the presence of P Cygni profiles 
in the $UV$ {\sl HST} spectra of
two other very metal-deficient galaxies, SBS 0335--052 ($Z_\odot$/40) and 
Tol 1214--277 ($Z_\odot$/23). 
In this paper we continue our search
for stellar populations with mass loss in very metal-deficient BCDs. We
present high signal-to-noise ratio optical spectrophotometry of I Zw 18. We 
show that WR stars of different types are 
clearly present in this galaxy.

   I Zw 18 is a BCD undergoing an intense burst of star formation. It was first 
recognized to have an exceptionally low metal abundance by
Searle \& Sargent (1972). Later studies by Lequeux et al. (1979), French
(1980), Kinman \& Davidson (1981), Pagel et al. (1992), Skillman \& 
Kennicutt (1993), Martin (1996), Izotov, Thuan \& Lipovetsky (1997)
and Izotov \& Thuan (1997b) have 
confirmed the oxygen abundance to be only $\sim$1/50 of the solar value.  
Zwicky (1966) described I Zw 18 as a double system of compact galaxies,
which are in fact two bright centers of star formation
separated by an angular distance of 5\farcs8. These two star-forming
regions will be referred to subsequently as
the brighter NW and fainter SE components. Later studies 
(Davidson, Kinman \& Friedman 1989; Dufour \& Hester 1990) have revealed a
more complex structure with several additional diffuse features.
However the NW and SE components dominate in brightness and this paper
will focus on them.
We describe the observations and
data reduction in \S2. In \S3 we discuss the properties of WR stars
in the BCD. We summarize our findings in \S4.

\section {OBSERVATIONS AND DATA REDUCTION}

   Spectrophotometric observations of I Zw 18
were obtained with the {\sl Multiple Mirror Telescope (MMT)} on the nights
of 1997 April 29 and 30. Observations were made with the blue channel of the 
{\sl MMT} spectrograph using a highly optimized Loral 3073$\times$1024 CCD
detector. A 1\farcs5$\times$180\arcsec\ slit was used along with a 300 groove
mm$^{-1}$ grating in first order and an L-38 second-order blocking filter. 
This yields a spatial scale along the slit of 0\farcs3 pixel$^{-1}$, a scale
perpendicular to the slit of 1.9\AA\ pixel$^{-1}$, a spectral range of
3600 -- 7500\AA, and a spectral resolution of $\sim$ 7\AA\ (FWHM). For these
observations, CCD rows were binned by a factor of 2, yielding a final sampling
of 0\farcs6 pixel$^{-1}$. The observations cover the full spectral range in a
single frame that contains all the lines of interest and have sufficient 
spectral resolution to distinguish between narrow nebular and broad WR emission
lines. The total exposure time was 180 minutes and was broken up into six
sub-exposures, 30 minutes each. All exposures were taken at small air masses
(1.1 -- 1.2), so no correction was made for atmospheric dispersion. The seeing
 was 0\farcs7 FWHM. The slit was oriented in the direction
with position angle P.A. = --41\arcdeg\ to permit observations of both NW
and SE components. The spectrophotometric standard stars EG 247 and HZ 44 were
observed for flux calibration. Spectra of He-Ne-Ar comparison lamps were
obtained after each subexposure to provide flux calibration.

     The two-dimensional spectra were bias subtracted and flat-field corrected
using the IRAF
\footnote[3]{IRAF is distributed by the National Optical Astronomy 
Observatories, which is operated by the Association of Universities for 
Research in Astronomy, Inc., under cooperative agreement with the National 
Science Foundation.}. For the NW component, the extracted one-dimensional 
spectra cover the brightest part of the galaxy 
with a spatial size of $\sim$5\arcsec. Similar procedures were used 
for the SE component, resulting in one-dimensional
spectra covering a region 5\arcsec\ wide at a distance of 5\farcs8 from 
the NW component.
The extracted spectra from each frame were then coadded 
and calibrated to absolute fluxes.

The observed line intensities have been corrected for interstellar extinction 
using the reddening law by Whitford (1958).
Hydrogen lines have been also corrected for underlying stellar absorption, with 
the equivalent width for hydrogen absorption lines derived self-consistently
together with the extinction coefficient from the observed intensities of all
hydrogen lines. We show the observed $F$($\lambda$)/$F$(H$\beta$) and 
extinction and absorption-corrected $I$($\lambda$)/$I$(H$\beta$) line 
intensities for the NW and SE components in Table 1, along with the extinction 
coefficient ($C$((H$\beta$)) and the equivalent width of the hydrogen absorption lines, the 
observed flux and equivalent width of the H$\beta$ emission line. The 
uncertainties for the tabulated relative line intensities are $\sim$0.5\% for
the strongest lines and $\sim$10 -- 15\% for the weakest lines.

\section{WOLF-RAYET STARS IN I ZW 18}


    In Figure 1 we show the spectra of the NW and the SE 
components of I Zw 18. The
WR broad lines and strong narrow He II $\lambda$4686 emission line
are clearly present in the spectrum of the NW component. 
Thus, I Zw 18 is the lowest metallicity galaxy where
WR stars are detected. In contrast, in the spectrum of the SE component
WR emission lines are not convincingly detected. The weak emission seen at the 
wavelength $\sim$ 4700 \AA\ is nebular emission from 
the [Fe III] $\lambda$4658, He II $\lambda$4686 and [Ar IV]  
$\lambda\lambda$4711, 4740 lines.  The appearance of a broad underlying emission
feature here most probably arises from the blending of the wings of the
individual nebular line
profiles, but a very small contribution from WR stars cannot be excluded.
The detection of the C IV $\lambda$5808 broad line in the NW component suggests 
the presence of early WC stars, while the other broad lines are, most likely, 
evidence of early and late WN stars. The formation of C IV $\lambda$5808
by WN stars can be ruled out because of its large FWHM ($\sim$80\AA) which
corresponds to early type WC stars (Smith, Shara \& Moffat 
1990).
The approximate number of O and WR stars in the NW component can be estimated
as follows. The observed fluxes of the C IV
$\lambda$5808 and broad He II $\lambda$4686 emission lines are 
8.3$\times$10$^{-16}$ and 1.49$\times$10$^{-15}$ erg s$^{-1}$cm$^{-2}$, and
that of H$\beta$ $\lambda$4861 integrated along the slit 
is 4.93$\times$10$^{-14}$ erg s$^{-1}$cm$^{-2}$.
Correcting for interstellar extinction $C$(H$\beta$) = 0.13 dex for the NW
component and adopting a distance $D$ = 10.8 Mpc (for the redshift 
$z$=0.00274 of the NW component with 
H$_0$ = 75 km s$^{-1}$ Mpc$^{-1}$), we derive the following
luminosities: $L$(C IV $\lambda$5808) = 1.27$\times$10$^{37}$ erg s$^{-1}$,
$L$(He II $\lambda$4686) = 2.84$\times$10$^{37}$ erg s$^{-1}$ and
$L$(H$\beta$) = 9.26$\times$10$^{38}$ erg s$^{-1}$. Assuming that only 1/2 of 
the light in H$\beta$ is contained within the slit width of 1\farcs5, 
 we derive finally $L$(H$\beta$) = 1.85$\times$10$^{39}$ erg s$^{-1}$. 
This value agrees well with that derived by Hunter \& Thronson (1995) from
the {\sl HST} H$\alpha$ image.
Adopting a value of $Q_0$ = 49.05 (Vacca \& Conti 1992) for the logarithm of 
the number of Lyman continuum
photons emitted per second  by an O7V star and assuming Case B 
recombination and an instantaneous burst of star formation,
we derive $N$(O7V) = 381. 
To derive the total number of O stars, we need to take into account the age of 
the stellar population and the
IMF slope. A measure of the age of the burst of star 
formation is the equivalent width of the H$\beta$ emission line. The low 
equivalent width
of 56\AA\ in the NW component of I Zw 18 corresponds to an age of 4 -- 5 Myr,
assuming a Salpeter IMF 
(Leitherer \& Heckman 1995). This value is in good agreement with the results 
of
direct {\sl HST} photometry of the brightest stars by Hunter \& Thronson 
(1995).
They found stars as old as 5 Myr in the NW component.
Then, the total number of O stars is $\sim$ 3
times larger (Schaerer 1996) and is equal to $\sim$1100. 
This approximate value is in good agreement with the number of O stars of 1300
derived by Hunter \& Thronson (1995) from the luminosity function.

	The number of WR stars is estimated from the luminosity of the WR emission
features.  While the main contribution to the $\lambda$5808 emission is from
WC4 stars, the emission in the $\lambda$4650 feature is produced by WR stars
of different types.  However, the observed flux ratio, f($\lambda$4650)/f($\lambda$5808)$\sim$3,
is significantly greater than that expected for WC4 stars (Smith, Shara, \& 
Moffat 1990).  Therefore, the dominant contributors to the broad He II
$\lambda$4686 emission are WNL stars.  The moderate spectral resolution and
line blending, however, prevent drawing more decisive conclusions about the
origin of the $\lambda$4650 bump.  The derived number of WNL stars should
therefore be considered as only indicative.
Adopting the luminosity of a single WC4 star 
in the C IV $\lambda$5808 line as equal 
to 2.5$\times$10$^{36}$ erg s$^{-1}$, and that of a single WNL star in He II 
$\lambda$4686 as equal to 1.7$\times$10$^{36}$ erg s$^{-1}$ (Conti 1991;
Vacca \& Conti 1992), we derive the following numbers of stars:
$N$(WC4) = 5$\pm$2, $N$(WNL) = 17$\pm$4. This gives $N$(WR)/$N$(O) = 0.02 
and $N$(WC)/$N$(WN) = 0.3.

     Meynet (1995) presented new evolutionary population synthesis models
based on the most recent grids of stellar models computed at the Geneva
Observatory. He studied the effects of changes in the rates of mass loss by
stellar winds on the massive star populations born in a
starburst. 
According to Maeder \& Meynet (1994), the high mass loss rate stellar
models are to be preferred over the standard ones on the basis of comparisons
with the observed luminosities, chemical compositions and number statistics
of WR stars in zones of constant star formation rate. In starburst galaxies,
the presence of WC stars at very low metallicity is predicted only by models of 
massive star evolution with enhanced mass loss (Meynet 1995), while the models 
with standard mass loss rates derived by de Jager, Nieuwenhuijzen \& van der 
Hucht (1988) and scaled with metallicity as $Z$$^{0.5}$
fail to produce WC stars at the metallicity of I Zw 18. Therefore, the
detection of WC stars in I Zw 18 gives strong support to the idea of enhanced
mass loss in massive low metallicity 
stars. Furthermore, we find satisfactory agreement 
between the observed and theoretical WR/O and WC/WN ratios at the metallicity
of I Zw 18. Meynet (1995) has calculated evolutionary population synthesis
models only for metallicities as low as $Z_\odot$/20. At this 
metallicity, assuming an instantaneous burst of star formation and an IMF
$dN/dM$$\propto$$M^{-2}$ for the massive stars, his models predict 
WR/O$\approx$0.035 and WC/WN$\approx$0.5 for peak values. Scaling 
these values as $Z^{0.5}$ to the I Zw 18 heavy element abundance 
$Z_\odot$/50 gives WR/O
and WC/WN ratios close to those derived from the observations.

     The detection of WC stars in I Zw 18 can help resolve the 
long-standing problem of the origin of the strong nebular He II $\lambda$4686 
line in the NW component, which is several orders
of magnitude greater than predicted by photoionized H II region models.
Schaerer (1996) has shown that WC stars can significantly increase the ionizing 
flux shortward of 228\AA\ , thus
leading to the formation of a He$^{++}$ zone in the
H II region and increasing the
recombination He II $\lambda$4686 emission line luminosity by several orders
of magnitude. However, this model predicts the maximum value of the 
nebular He II $\lambda$4686 emission line intensity to be only
$\sim$1\% -- 2\% that
of H$\beta$ when the age of the star forming region is not greater than 3 Myr.
The He II line intensity is expected to decrease from this maximum value for a
somewhat older stellar population like the one in I Zw 18. This is in contrast 
to 
the observed intensity of the nebular He II $\lambda$4686 emission 
line in the NW component of I Zw 18 which is 4\% that of H$\beta$, larger than 
the predicted peak value. Furthermore, this
model fails to explain the presence of strong
He II $\lambda$4686 in the galaxies where WC stars have not been detected
(Izotov et al. 1997). We conclude therefore that, in addition to WC stars, some 
other source of
ionizing radiation at wavelengths shorter than 228\AA\ must be
invoked.

\section{CONCLUSIONS}

   We have obtained high signal-to-noise ratio spectrophotometric observations 
of I Zw 18, the most metal-deficient BCD known, in an attempt to detect
the broad low-intensity emission lines of WR stars. We have obtained 
the following results:

    1. Broad emission lines at wavelengths $\sim$ $\lambda$4650 and 
$\lambda$5800 have been detected in the spectrum of the NW component 
of I Zw 18 implying the presence of WN and WC stars, while in the 
younger SE component these lines are absent. Thus, I Zw 18 is the 
lowest-metallicity Wolf-Rayet galaxy known to-date. 

    2. The total numbers of WNL and WC4 stars in I Zw 18 are 17$\pm$4 and 5$\pm$2 
respectively, and the total number of O stars is $\sim$ 1100. The existence
of WC stars at the very low metallicity of $Z_\odot$/50 in I Zw 18 confirms
predictions of massive stellar evolution models with enhanced mass loss
rates (Meynet 1995), while the models with standard mass loss rates fail to
explain the presence of WC stars in I Zw 18.
The observed WR/O and WC/WN ratios of 0.02 and 0.3 respectively, 
are in satisfactory agreement with evolutionary population synthesis
models based on stellar evolution models with enhanced mass loss and 
extrapolated to the metallicity of I Zw 18.
Although the WC stars detected in I Zw 18 could be responsible
for the presence of the strong nebular He II $\lambda$4686 emission line,
the observed value of the nebular He II $\lambda$4686, 4\% that of
H$\beta$, is several times larger than theoretical predictions. We conclude
that additional sources of ionizing radiation at wavelengths shorter
than 228\AA\ need to be present.

\acknowledgements
This international collaboration has been made possible by the support of
INTAS research grant No 94-2285 and NATO collaborative research grant 921285 . 
Y.I.I. is grateful for the hospitality of the 
National Optical Astronomy Observatories and the MMT Observatory. C.B.F.
acknowledges the support of the NSF through grant AST 93-20715. We thank
Phil Massey for useful discussions and the referee, Cesar Esteban, for
thoughtful comments that improved the content and presentation.

\clearpage

\figcaption[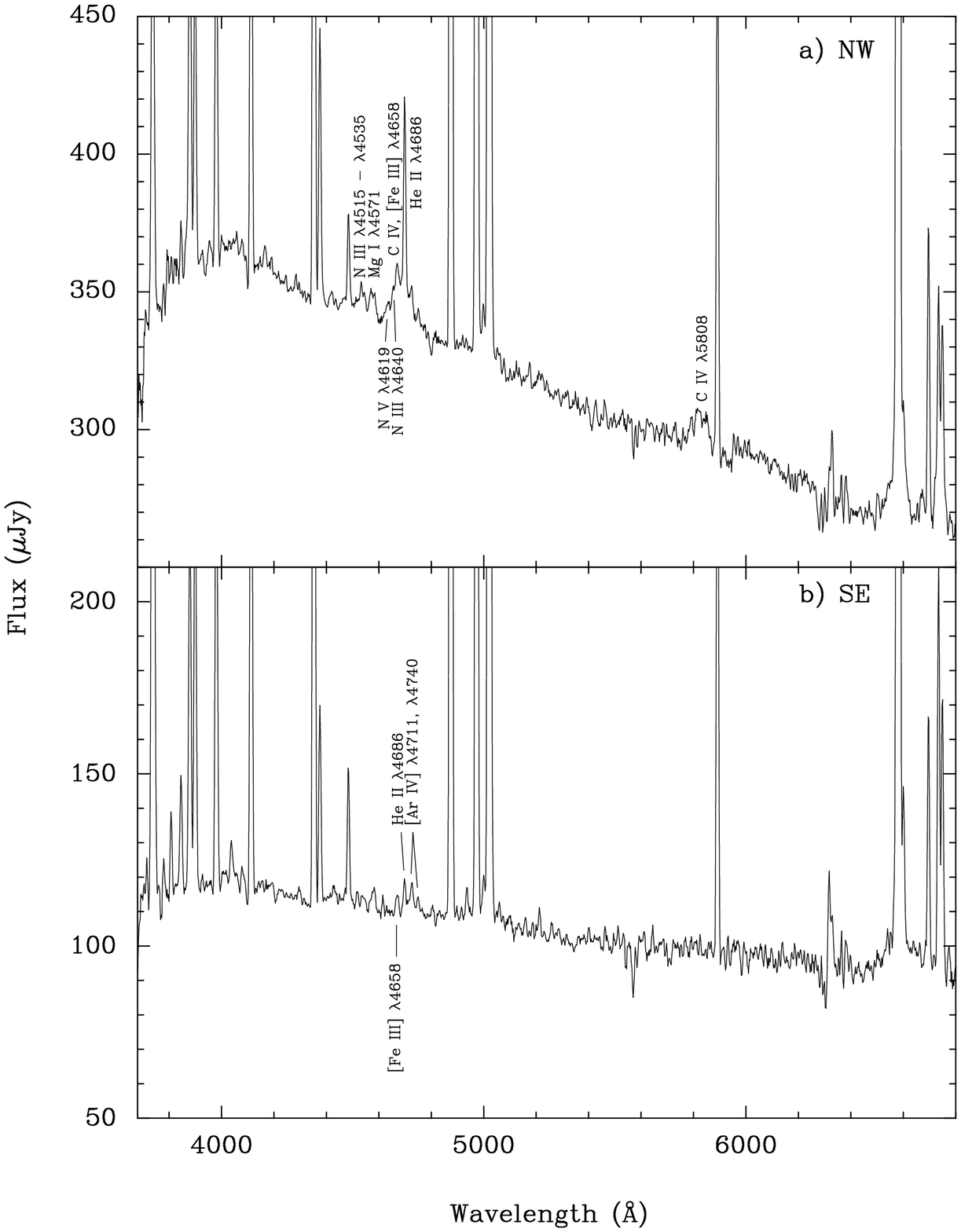]{The spectra of
the NW and the SE components in the BCD I Zw 18. The broad nitrogen
and carbon lines in the spectrum of the NW component
are marked indicating the presence of WN and
WC stars. The permitted Mg I $\lambda$4571 and narrow nebular
He II $\lambda$4686 emission lines are also marked. The broad
WR emission lines are absent in the spectrum of the SE component where
the nebular [Fe III] $\lambda$4658, He II $\lambda$4686 and [Ar IV]
$\lambda$$\lambda$4711, 4740 emission lines are marked.
The spectra have been smoothed using a 3-point box-car. 
\label{fig1}}

\begin{deluxetable}{lccccc}
\tablenum{1}
\tablecolumns{6}
\tablewidth{40pc}
\tablecaption{Emission Line Intensities}
\tablehead{
\colhead{} & \multicolumn{2}{c}{I Zw 18 (NW)} && \multicolumn{2}{c}{I Zw 18 
(SE)} \\ \cline{2-3} \cline{5-6} \\
\colhead{Ion}&\colhead{$F$($\lambda$)/$F$(H$\beta$)}
&\colhead{$I$($\lambda$)/$I$(H$\beta$)}&\colhead{}
&\colhead{$F$($\lambda$)/$F$(H$\beta$)}
&\colhead{$I$($\lambda$)/$I$(H$\beta$)} }
\startdata
 3727\ [O II]        & 0.228& 0.238& & 0.510& 0.499 \nl
 3868\ [Ne III]      & 0.143& 0.148& & 0.139& 0.136 \nl
 3889\ He I + H8     & 0.102& 0.195& & 0.155& 0.200 \nl
 3968\ [Ne III] + H7 & 0.134& 0.222& & 0.169& 0.212 \nl
 4101\ H$\delta$     & 0.196& 0.276& & 0.232& 0.270 \nl
 4340\ H$\gamma$     & 0.410& 0.472& & 0.443& 0.468 \nl
 4363\ [O III]       & 0.069& 0.068& & 0.054& 0.052 \nl
 4471\ He I          & 0.022& 0.021& & 0.034& 0.033 \nl
 4686\ He II (neb)   & 0.041& 0.040& & 0.007& 0.006 \nl
 4686\ He II (WR)    & 0.063& 0.062& & \nodata& \nodata \nl
 4861\ H$\beta$      & 1.000& 1.000& & 1.000& 1.000 \nl
 4959\ [O III]       & 0.731& 0.690& & 0.591& 0.573 \nl
 5007\ [O III]       & 2.197& 2.069& & 1.776& 1.724 \nl
 5808\ C IV (WR)     & 0.035& 0.031& & \nodata& \nodata \nl
 5876\ He I          & 0.074& 0.066& & 0.095& 0.092 \nl
 6563\ H$\alpha$     & 3.163& 2.743& & 2.834& 2.745 \nl
 6678\ He I          & 0.030& 0.026& & 0.028& 0.027 \nl
 6717\ [S II]        & 0.024& 0.021& & 0.043& 0.042 \nl
 6731\ [S II]        & 0.019& 0.016& & 0.031& 0.030 \nl
 7065\ He I          & 0.026& 0.022& & 0.024& 0.024 \nl
 7135\ [Ar III]      & 0.019& 0.016& & 0.019& 0.018 \nl \nl
 $C$(H$\beta$) dex    &\multicolumn {2}{c}{0.130}&\colhead{}&\multicolumn 
{2}{c}{0.010} \nl
 $F$(H$\beta$)\tablenotemark{a} &\multicolumn {2}{c}{ 2.36}&\colhead{} 
&\multicolumn{2}{c}{1.76} \nl
 $EW$(H$\beta$)\ \AA &\multicolumn {2}{c}{ 56}&&\multicolumn {2}{c}{129} \nl
 $EW$(abs)\ \AA      &\multicolumn {2}{c}{2.9}
&\colhead{}&\multicolumn{2}{c}{3.9} \nl
\enddata

\tablenotetext{a}{in units of 10$^{-14}$ ergs\ s$^{-1}$cm$^{-2}$}

\end{deluxetable}




\clearpage

\begin{figure*}
\figurenum{1}
\epsscale{2.0}
\plotfiddle{fig1.ps}{0.cm}{0.}{100.}{100.}{-320.}{-320.}
\end{figure*}

\end{document}